\documentclass[12pt]{article}

\usepackage{a4wide}
\usepackage[pdftex,usenames,dvipsnames]{color}
\usepackage{graphics}
\usepackage{amsfonts}

\newcommand{\nit}{\noindent}

\newcommand{\np}{\newpage}
\newcommand{\dsp}{\displaystyle}
\newcommand{\vs}[1]{\vspace{#1 ex}}
\newcommand{\hs}[1]{\hspace{#1 em}}
\newcommand{\bfr}{\begin{flushright}}
\newcommand{\efr}{\end{flushright}}
\newcommand{\bc}{\begin{center}}
\newcommand{\ec}{\end{center}}
\newcommand{\ben}{\begin{enumerate}}
\newcommand{\een}{\end{enumerate}}

\newcommand{\be}{\begin{equation}}
\newcommand{\ee}{\end{equation}}
\newcommand{\ba}{\begin{array}}
\newcommand{\ea}{\end{array}}
\newcommand{\ct}{\cite}
\newcommand{\bit}{\bibitem}
\newcommand{\dd}[2]{\frac{\partial{#1}}{\partial{#2}}}

\newcommand{\bg}{\beta}

\newcommand{\del}{\delta}

\newcommand{\ve}{\varepsilon}

\newcommand{\thg}{\theta}
\newcommand{\kg}{\kappa}
\newcommand{\lb}{\lambda}
\newcommand{\sg}{\sigma}
\newcommand{\rg}{\rho}

\newcommand{\fg}{\phi}
\newcommand{\vf}{\varphi}
\newcommand{\og}{\omega}

\newcommand{\Del}{\Delta}

\newcommand{\Og}{\Omega}

\newcommand{\bfh}{\bold{h}}

\newcommand{\bfer}{\bold{r}}

\newcommand{\bfx}{\bold{x}}

\newcommand{\bfE}{\bold {E}}

\newcommand{\bfI}{\bold {I}}
\newcommand{\bfJ}{\bold {J}}
\newcommand{\bfL}{\bold {L}}
\newcommand{\bfM}{\bold {M}}
\newcommand{\bfN}{\bold {N}}
\newcommand{\bfQ}{\bold {Q}}
\newcommand{\bfR}{\bold {R}}

\newcommand{\uh}{\underline{h}}

\newcommand{\cE}{{\cal E}}

\newcommand{\cH}{{\cal H}}
\newcommand{\cJ}{{\cal J}}

\newcommand{\cM}{{\cal M}}

\newcommand{\cP}{{\cal P}}
\newcommand{\cS}{{\cal S}}

\newcommand{\sfr}{{\sf r}}

\newcommand{\lh}{\left(}
\newcommand{\rh}{\right)}
\newcommand{\ld}{\left.}
\newcommand{\rd}{\right.}

\newcommand{\ddd}[1]{\stackrel{\cdots}{#1}}

\newcommand{\der}{\partial}

\begin{document}

\pagestyle{empty}

\bc
{\Large \bf Gravitational waves}\\
\vs{2}

{\Large \bf from generalized newtonian sources} 
\vs{7} 

{\large J.W.\ van Holten}
\vs{5}

{\large Lorentz Institute}
\vs{1}

{\large Leiden University, Leiden NL}
\vs{2} 

and
\vs{2} 

{\large Nikhef, Amsterdam NL}
\vs{5}

February 10, 2019
\ec
\vs{7}

\nit
{\small 
{\bf Abstract} \\
I review the elementary theory of gravitational waves on a Minkowski background and the quadrupole 
approximation. The modified conservation laws for energy and momentum keeping track of the 
gravitational-wave flux are presented. The theory is applied to two-body systems in bound and 
scattering states subject to newtonian gravity generalized to include a $1/r^3$ force allowing 
for orbital precession. The evolution of the orbits is studied in the adiabatic approximation. 
From these results I derive the conditions for capture of two bodies to form a bound state by 
the emission of gravitational radiation.
}

\np
\pagestyle{plain}
\pagenumbering{arabic}

\section{Introduction and overview \label{s1}}

The existence of gravitational waves is now well-established from both direct and indirect observations
\ct{Taylor1989}-\ct{LIGO2017b}.  A completely new field of astronomy is opening up which will no doubt 
have an impact also on other branches of astronomy and astrophysics such as dynamics and evolution 
of stars and galaxies. The supermassive black holes in the centers of galaxies, and possibly 
intermediate-mass black holes in stellar clusters, will by the relatively large curvature they create in the 
surrounding space enhance the emission of gravitational waves from massive objects on trajectories 
passing close to them, whether these are on bound or open orbits. The emission of gravitational waves 
can even lead to the capture of objects originally in open orbits to end up in a bound state. 

Apart from these radiative phenomena involving very massive black holes, the emission of gravitational 
waves also affects more common binary star systems like the well-known close binary neutron stars, 
the recently discovered binary black holes and presumably systems containing white dwarfs 
\ct{Nelemans2018}. No doubt radiation has an impact on three- and many-body systems, 
especially on their stability. Detailed investigations of close binary star systems using high-order 
post-newtonian expansions of the Einstein equations of General Relativity have been carried out 
with great success; for a review see e.g.\ \ct{Blanchet2014}. The inspiral and merger of extreme 
mass-ratio binaries involving a very massive black hole has also been studied directly in the 
background geometry of the black hole \ct{Lousto1997}-\ct{Dambrosi2014}. Whenever these 
theoretical investigations can be compared with data they seem to describe the dynamics of these 
systems very well, thereby also confirming General Relativity to be the best available theory for 
gravitational interactions \ct{LIGO2016b}. The study of radiation from two-body scattering 
has been addressed as well \ct{Damour1981}, although no corresponding observations have been 
announced so far.
 
Even though they may carry large amounts of energy and momentum, the deformations of space-time 
created by gravitational waves are extremely small. For example a flux of monochromatic gravitational 
waves with a frequency of 100 Hz and an extreme intensity of 1 W/m$^2$ will create spatial 
deformations of less than 1 part in $10^{19}$, the diameter of a proton over a distance of 1 km. 
This testifies as to the  extreme stiffness of space and explains both why it is so difficult to create 
gravitational waves and to observe them. It also implies that most potential sources of gravitational 
waves are weak and many move on close-to-stationary almost-newtonian orbits. 

This review is devoted to gravitational radiation from such weak or very weak sources. They produce 
the most abundant, though maybe not the most spectacular, form of gravitational waves in the 
universe and may eventually become relevant to a wide range of astronomical and astrophysical 
observations. To lowest order their description and propagation involve straightforward applications 
of linear field theory in Minkowski space-time. This also provides the starting point for many more 
elaborate and precise calculations.

We will begin by recapturing in fairly standard fashion the wave equation for gravitational waves, its 
gauge invariance and its implications for the propagation and polarization states of gravitational waves. 
We address the quadrupole nature of the waves and the associated sources, and explain how dynamical 
mass quadrupole motion generates the simplest and most common weak gravitational waves. Next we 
derive the modification of the conservation laws for energy, momentum and angular momentum by taking 
account of gravitational radiation. We present equations for the transport of energy and angular momentum 
by gravitational waves, keeping track of the anisotropic dependence on directions.

This theory is then applied to systems of massive objects moving on generalized newtonian orbits, either in 
bound states or on open scattering trajectories. The generalization includes the effects of possible $1/r^3$ 
forces causing orbital precession, which may result e.g.\ from many-body or post-newtonian interactions. 
We calculate the evolution of orbital parameters due to emission of gravitational radiation and their relations. 
We finish by establishing which binary scattering orbits are turned into bound states by emission of radiation.

\section{The wave equation \label{s2}}

Weak gravitational waves are dynamical fluctuations of the space-time metric about flat Minkowski 
geometry \ct{Einstein1918,MTW,Maggiore2008}. Thus we can split the full space-time metric as 
\be
g_{\mu\nu} = \eta_{\mu\nu} + 2 \kg h_{\mu\nu},
\label{2.1}
\ee
where $\kg$ is the positive root of 
\be
\kg^2 = \frac{8 \pi G}{c^4} \simeq 2.1 \times 10^{-41}\, \mbox{kg$^{-1}$ m$^{-1}$ s$^2$},
\label{2.2}
\ee
$G$ being the newtonian constant of gravity and $c$ the speed of light in vacuum. This endows 
 $h_{\mu\nu}$ with the standard dimensions of a bosonic tensor field. Up to non-linear 
corrections the tensor field is postulated to satisfy the field equation 
\be
\Box h_{\mu\nu} - \der_{\mu} \der^{\lb} h_{\lb\nu} - \der_{\nu} \der^{\lb} h_{\lb\mu} + 
  \der_{\mu} \der_{\nu} h^{\lb}_{\;\,\lb} - \eta_{\mu\nu} \lh \Box h^{\lb}_{\;\,\lb} 
  - \der^{\kg} \der^{\lb} h_{\kg\lb} \rh = - \kg T_{\mu\nu},
\label{2.3}
\ee
where $\Box = \eta^{\mu\nu} \der_{\mu} \der_{\nu}$ is the d'Alembertian and  the inhomogeneous 
term $T_{\mu\nu}$ on the right-hand side represents the sources of the field. By factoring out the 
constant $\kg$ this tensor has the dimensions of energy per unit of volume or force per unit of area. 
In this treatise we always use the flat Minkowski metric $\eta_{\mu\nu}$ with signature $(-,+,+,+)$ 
and its inverse $\eta^{\mu\nu}$ to raise and lower indices on components of mathematical objects 
like vectors and tensors. 

The motivation for postulating this field equation comes from the physical properties of the tensor 
field $h_{\mu\nu}$ implied by its structure. First note that defining the linear Ricci tensor
\be
R_{\mu\nu} = \kg \lh \Box h_{\mu\nu} - \der_{\mu} \der^{\lb} h_{\lb\nu} - \der_{\nu} \der^{\lb} h_{\lb\mu} + 
  \der_{\mu} \der_{\nu} h^{\lb}_{\;\,\lb} \rh,
\label{2.4}
\ee
the trace of which reads
\be
R = R^{\lb}_{\;\,\lb} = 2 \kg \lh \Box h^{\lb}_{\;\,\lb} - \der^{\kg} \der^{\lb} h_{\kg\lb} \rh,
\label{2.5}
\ee
the field equation takes the form
\be
R_{\mu\nu} - \frac{1}{2}\, \eta_{\mu\nu} R = - \kg^2\, T_{\mu\nu}.
\label{2.6}
\ee
This is the linearized version of Einstein's gravitational field equation in a flat background. 
Note also that 
\be
\der^{\mu} R_{\mu\nu} = \frac{1}{2}\, \der_{\nu} R,
\label{2.6a}
\ee
and as a result the inhomogeneous field equation (\ref{2.6}) is seen to imply a conservation law 
for the source terms:
\be
\der^{\mu} T_{\mu\nu} = 0.
\label{2.7}
\ee
As the energy-momentum tensor of matter and radiation has the required physical dimensions 
and satisfies the condition (\ref{2.7}) in Minkowski space it is the obvious source for the tensor 
field. As all physical systems possess energy and momemtum this explains the universality of 
gravity\footnote{As is well-known, requiring this universality to encompass the gravitational field 
itself leads to the non-linear structure of the full theory of General Relativity.}.

An observation closely related to (\ref{2.6a}) is that the linear Ricci tensor is invariant under gauge
transformations
\be
h_{\mu\nu}\; \rightarrow\; h'_{\mu\nu} = h_{\mu\nu} + \der_{\mu} \xi_{\nu} + \der_{\nu} \xi_{\mu}, \hs{2} 
R'_{\mu\nu} = R_{\mu\nu}.
\label{2.8}
\ee
By such gauge transformations one can straightforwardly eliminate four components of the field to 
reduce the number of independent components from ten to six. To achieve such a reduction in 
practice the standard procedure is to impose the De Donder condition
\be
\der^{\mu} h_{\mu\nu} = \frac{1}{2}\, \der_{\nu} h^{\mu}_{\;\,\mu}.
\label{2.9}
\ee
This condition reduces the linear Ricci tensor and its trace to the expressions
\be
R_{\mu\nu} = \kg\, \Box h_{\mu\nu}, \hs{2} R = \kg\, \Box h^{\lb}_{\;\,\lb},
\label{2.10}
\ee
and therefore the field equation turns into the inhomogeneous wave equation
\be
\Box \lh h_{\mu\nu} - \frac{1}{2}\, \eta_{\mu\nu} h^{\lb}_{\;\,\lb} \rh = - \kg T_{\mu\nu}.
\label{2.11}
\ee
It is then convenient to redefine the field components by
\be
\uh_{\mu\nu} \equiv h_{\mu\nu} - \frac{1}{2}\, \eta_{\mu\nu} h^{\lb}_{\;\,\lb},
\label{2.12}
\ee
which transform under gauge transformations as 
\be
\uh'_{\mu\nu} = \uh_{\mu\nu} + \der_{\mu} \xi_{\nu} + \der_{\nu} \xi_{\mu} - \eta_{\mu\nu}\, \der^{\lb} \xi_{\lb}.
\label{2.13}
\ee
After implementing the De Donder condition the field is divergence-free and satisfies the 
inhomogeneous wave equation:
\be
\der^{\mu} \uh_{\mu\nu} = 0, \hs{2} \Box \uh_{\mu\nu} = - \kg T_{\mu\nu}.
\label{2.14}
\ee
Finally a second gauge transformation can be made without changing the De Donder condition 
provided the parameter satisfies itself the homogeneous wave equation: 
\be
\der^{\mu} \uh'_{\mu\nu} = \der^{\mu} \uh_{\mu\nu} + \Box\, \xi_{\nu} = 0 \hs{1} \Leftrightarrow \hs{1} 
\Box\, \xi_{\nu} = 0.
\label{2.15}
\ee
Such a residual gauge transformation can be made in particular on free fields to remove the 
trace of the tensor field:
\be
\uh^{\prime\,\lb}_{\;\;\lb} = \uh^{\lb}_{\;\,\lb} - 2\, \der^{\lb} \xi_{\lb} = 0,
\label{2.16}
\ee
in agreement with the equations (\ref{2.14}) provided $\Box \xi_{\nu} = 0$ and $T^{\lb}_{\;\,\lb} = 0$.
It follows automatically that the same condition holds for the original tensor field: $h^{\lb}_{\;\,\lb} = 0$.
Removal of the trace reduces the number of independent components of free fields to five, equal to 
the dimension of the irreducible spin-2 representation of the rotation group, However, as dynamical 
free wave fields propagate on the light cone and have only transverse polarization states, the actual 
number of independent {\em dynamical} components of gravitational wave fields is two. This will be 
discussed in the following.

\section{Solutions of the inhomogeneous wave equation \label{s3}} 

The inhomogeneous linear wave equation (\ref{2.14}) has many solutions: to a given solution one 
can always add any solution of the homogeneous equation representing free gravitational waves. 
Free gravitational waves can therefore appear as a background to gravitational wave signals from 
specific sources. 

In the absence of such a background the standard causal solution for sources localized in a finite 
region of space is the retarded solution  
\be
\uh_{\mu\nu}(\bfx,t) = \frac{\kg}{4\pi}\, \int_{S_{\sfr}} d^3x'\, 
  \frac{T_{\mu\nu}(\bfx', t - |\bfx' - \bfx|)}{|\bfx' - \bfx|},
\label{3.1}
\ee
where the integration volume $S_{\sfr}$ can be taken to be a large sphere of radius $\sfr = |\bfx|$ 
containing the finite region of the sources where $T_{\mu\nu} \neq 0$ in its center. To evaluate the 
field by performing the integration is difficult in practice for any realistic type of sources. 

In order to make progress it makes sense to consider the situation in which the waves are evaluated 
at large distance from the sources: the radius $\sfr$ of the sphere is taken to be much larger than 
any typical dimension of the sources. For example we evaluate the waves emitted by a binary star 
system of orbital extension $d$ at a distance $\sfr \gg d$. Under this assumption one can expand the
integral expression on the right-hand side of (\ref{3.1}) in inverse powers of $\sfr$ keeping only terms 
which do not fall off faster than $1/\sfr$. This results in the simpler integral
\be
\uh_{\mu\nu}(\bfx,t) = \frac{\kg}{4\pi \sfr}\, \int_{S_{\sfr}} d^3x'\, T_{\mu\nu}(\bfx', t-\sfr).
\label{3.2}
\ee
Another simplification is possible as it is straightforward to show that for localized sources 
these solutions have no dynamical time components:
\be
\ba{lll}
\der_0 \uh_{0\mu} & = & \dsp{ \frac{\kg}{4\pi \sfr}\, \int_{S_{\sfr}} d^3 x'\, \der_0 T_{0\mu} 
  = \frac{\kg}{4\pi \sfr}\, \int_{S_{\sfr}} d^3x'\, \der'_i\, T_{i\mu} }\\
 & & \\
 & = & \dsp{ \frac{\kg}{4\pi \sfr}\, \oint_{\der S_{\sfr}} d^2 \sg\, \hat{\sfr}'_i\, T_{i\mu} = 0. }
\ea
\label{3.3}
\ee
The second equality on the first line follows from energy-momentum conservation, whilst 
the last equality uses Gauss' theorem to convert the volume integral to a surface integral 
over the corresponding normal component of the energy-momentum tensor, $\hat{\sfr}$ 
being the radial unit vector pointing out of the spherical surface $\der S_{\sfr}$. Finally the 
localization of the sources in a finite region near the center of the sphere guarantee the 
vanishing of the energy-momentum tensor on the boundary. We infer that the time 
components may represent static newtonian fields, but they cannot contribute to the flux 
of dynamical waves across the boundary of the sphere. 

As concerns dynamical fields we are therefore left with the spatial components of the 
outgoing wave solutions (\ref{3.2}):
\be
\uh_{ij} = \frac{\kg}{4\pi \sfr}\, \int_{S_{\sfr}} d^3x'\, T_{ij}(\bfx', t-\sfr).
\label{3.4}
\ee
In empty space far from the sources the expression on the right-hand side actually 
represents an exact formal solution of the wave equation. Now this solution was 
obtained by imposing the De Donder condition (\ref{2.14}); in addition, as argued after
(\ref{2.16}), in this region one can always find a local gauge transformation of the fields 
that makes them traceless. For the solution at hand this implies that after such a gauge 
transformation 
\be
\der_i \uh_{ij} = 0 \hs{1} \Rightarrow \hs{1} \hat{\sfr}_i\, \uh_{ij} = 0.
\label{3.5}
\ee
and
\be
\uh_{jj} = h_{jj} = 0.
\label{3.6}
\ee
A detailed discussion of the necessary gauge transformations is presented in appendix
\ref{a0}. Tensor fields obeying these conditions of are called transverse and traceless $(TT)$ 
and satisfy $\uh_{ij}^{TT} = h_{ij}^{TT}$.  We will take these properties for granted in what 
follows and omit the $TT$ in the notation. Combining the above requirements the outgoing 
wave fields far from the source must then be represented in the $TT$-gauge by an expression 
of the form
\be
h_{ij}(\bfx,t) = \uh_{ij}(\bfx,t) = \frac{\kg}{4\pi \sfr}\, \lh \del_{ik} - \hat{\sfr}_i \hat{\sfr}_k \rh \lh \del_{jl} - 
 \hat{\sfr}_j \hat{\sfr}_l \rh \lh I_{kl} + \frac{1}{2}\, \del_{kl}\, \hat{\sfr} \cdot I \cdot \hat{\sfr} \rh,
\label{3.7}
\ee
where the spatial symmetric 3-tensor $\bfI$ is traceless: $I_{kk} = 0$. Writing $u \equiv t - \sfr$,
agreement of this expression with the result (\ref{3.4}) up to gauge transformations is 
obtained by taking 
\be
I_{ij}(u) = \int_{S_{\sfr}} d^3 x'\, \lh T_{ij} - \frac{1}{3}\, \del_{ij} T_{kk} \rh \lh \bfx', u \rh.
\label{3.8}
\ee
With the help of energy-momentum conservation the integral can be rewritten in terms of the 
quadrupole moment of the total energy density $T_{00}$ of the sources:
\be
I_{ij}(u) = \frac{1}{2}\, \der_0^2 \int_{S_{\sfr}} d^3x' \lh x'_i x'_j  - \frac{1}{3}\, \del_{ij}\, 
 \bfx^{\prime\,2} \rh T_{00}(\bfx',u).
\label{3.9}
\ee
The proof is easier in backward fashion; first notice that as $\der_0 = \der_u$
\be
\der_0^2 T_{00}(\bfx',u) = \der_0\, \der'_i\, T_{i0} = \der'_i \der'_j\, T_{ij}(\bfx',u);
\label{3.10}
\ee
then perform two partial integrations with respect to $\bfx'$ to reobtain (\ref{3.8}), observing 
that the full energy-momentum tensor is supposed to vanish at the boundary $\der S_{\sfr}$. 

Finally considering non-relativistic sources in the center-of-mass frame, the energy density 
is dominated by the mass-density $\rg(\bfx,t)$, which allows us to replace the integral in 
(\ref{3.9}) by the components of the mass quadrupole moment and write explicitly:
\be
I_{ij} = \frac{1}{2}\, \frac{d^2 Q_{ij}}{dt^2}, \hs{2} 
Q_{ij}(u) = \int_{S_{\sfr}} d^3x' \lh x'_i x'_j - \frac{1}{3}\, \del_{ij} \, \bfx^{\prime\,2} \rh \rg(\bfx',u).
\label{3.11}
\ee
Thus we get the final expression for the wave field $\uh_{ij}$ for non-relativistic sources 
in the $TT$-gauge:
\be
h_{ij}(\bfx,t) = \frac{\kg}{8\pi \sfr}\, \lh \del_{ik} - \hat{\sfr}_i \hat{\sfr}_k \rh 
 \lh \del_{jl} - \hat{\sfr}_j \hat{\sfr}_l \rh
 \frac{d^2}{dt^2} \lh Q_{kl} + \frac{1}{2}\, \del_{kl}\, \hat{\sfr} \cdot Q \cdot \hat{\sfr}  \rh_{u = t-\sfr}.
\label{3.12}
\ee
For the dynamical (non-Newtonian) metric fluctuations $\del g_{\mu\nu} = g_{\mu\nu} - \eta_{\mu\nu}$, 
recalling equations (\ref{2.1}) and (\ref{2.2}) this result implies that 
\be
\ba{l} \del g_{00} = \del g_{0i} = 0;  \\
 \\
\dsp{ \del g_{ij} = \frac{2G}{\sfr} \lh \del_{ik} - \hat{\sfr}_i \hat{\sfr}_k \rh  \lh \del_{jl} - \hat{\sfr}_j \hat{\sfr}_l \rh 
 \dd{^2}{t^2} \lh Q_{kl} + \frac{1}{2}\, \del_{kl}\, \hat{\sfr} \cdot Q \cdot \hat{\sfr} \rh_{u = t-\sf r}. } 
\ea 
\label{3.13}
\ee

\section{Conservation laws and gravitational-wave fluxes \label{s4}}

Free radiation fields (always taken in the $TT$-gauge) define conserved currents of energy, 
momentum and angular momentum \ct{MTW, Maggiore2008}; in the conventions of the previous
sections
\be
\ba{l}
\dsp{ \cE = \frac{1}{2}\, \lh \der_0 h_{ij} \rh^2 + \frac{1}{2} \lh \der_k h_{ij} \rh^2, \hs{2} 
\cP_k = \der_0 h_{ij} \der_k h_{ij}, }\\
 \\
\cM_k = \der_0 h_{ij} \lh 2 \ve_{kmi} h_{mj} - \ve_{kmn} x_m \der_n h_{ij} \rh.
\ea
\label{4.1}
\ee
Subject to the field equations and gauge conditions these quantities satisfy the continuity equations 
\be
\dd{\cE}{t} = \der_j \cP_j, \hs{2} \dd{\cP_k}{t} = \der_j \cS_{jk}, \hs{2} 
\dd{\cM_k}{t} = \der_j \cJ_{jk},
\label{4.2}
\ee
where
\be
\ba{l}
\dsp{ \cS_{jk} = \der_j h_{mn} \der_k h_{mn} + \frac{1}{2}\, \del_{jk} 
 \left[ \lh \der_0 h_{mn} \rh^2 - \lh \der_l h_{mn} \rh^2 \right], }\\
 \\
\dsp{ \cJ_{jk} = 2 \ve_{kmn} h_{ml} \der_j h_{nl} - 
 \frac{1}{2}\, \ve_{jkl} x_l \left[ \lh \der_0 h_{mn} \rh^2 - \lh \der_l h_{mn} \rh^2 \right]. }
\ea
\label{4.3}
\ee
Applying them to the free fields (\ref{3.12}) these expressions determine the flux of energy, momentum 
and angular momentum carried by outgoing gravitational waves far from the source region. First, 
integration over a large sphere around the center of mass of the source and using Gauss' theorem 
gives the change in total energy, momentum and angular momentum of gravitational waves in 
terms of surface integrals
\be
\ba{l}
\dsp{ \frac{dE}{dt} = \oint_{\der S_{\sfr}} d^2 \sg\, \hat{\sfr}_i \cP_i, \hs{2} 
 \frac{dP_k}{dt} = \oint_{\der S_{\sfr}} d^2 \sg\, \hat{\sfr}_i \cS_{ik}, }\\
 \\
\dsp{ \frac{dM_k}{dt} = \oint_{\der S_{\sfr}} d^2 \sg\, \hat{\sfr}_i \cJ_{ik}. }
\ea
\label{4.4}
\ee
Next, on the spherical surface $\der S_{\sfr}$ the surface element of integration taken in polar 
co-ordinates $(\sfr, \thg, \vf)$ is
\be
d^2 \sg = \sfr^2 \sin \thg\, d\thg d \vf \equiv \sfr^2 d^2 \Og.
\label{4.5}
\ee
Evaluating the integrands on the right-hand side in equations (\ref{4.4}) while restoring factors of $c$ 
then results in differential fluxes
\be
\ba{l}
\dsp{ \frac{dE}{d^2 \Og dt} = - \frac{G}{8\pi c^5} \left[ \mbox{Tr}\, {\ddd{Q}} {\hs{0.1}^2} -
 2 \hat{\sfr} \cdot {\ddd{Q}}\,{^{2}} \cdot \hat{\sfr} + \frac{1}{2} (\hat{\sfr} \cdot \ddd{Q} \cdot \hat{\sfr}){^2} 
 \right]_{u = t - \sfr}, }\\
 \\
\dsp{ \frac{dP_k}{d^2 \Og dt} = - \frac{dE}{d^2 \Og\, c dt}\, \hat{\sfr}_k 
 = \frac{G}{8\pi c^6}\, \hat{\sfr}_k \left[ \mbox{Tr}\, {\ddd{Q}} {\hs{0.1}^2} -
 2 \hat{\sfr} \cdot {\ddd{Q}}\,{^{2}} \cdot \hat{\sfr} + \frac{1}{2} (\hat{\sfr} \cdot \ddd{Q} \cdot \hat{\sfr}){^2} 
 \right]_{u = t - \sfr}, }\\
 \\
\dsp{ \frac{dM_k}{d^2 \Og dt} = - \frac{G}{4\pi c^5}\, \ve_{kij} \left[ 
 \lh \ddot{Q} \cdot \ddd{Q} \rh_{ij} - \lh \ddot{Q} \cdot \hat{\sfr} \rh_i \lh \ddd{Q} \cdot \hat{\sfr} \rh_j \rd }\\
 \\
\dsp{ \hs{12} \ld +\, \hat{\sfr}_i \lh \ddot{Q} \cdot \ddd{Q} \cdot \hat{\sfr} - 
  \frac{1}{2}\, \ddot{Q} \cdot \hat{\sfr}\, \hat{r} \cdot \ddd{Q} \cdot \hat{\sfr} \rh_j \right]_{u = t - \sfr}. }
\ea
\label{4.6}
\ee
As usual overdots denote derivatives with respect to time $t$. The integrands themselves represent
the anisotropic angular distribution of fluxes. The spherical surface integrals can be performed 
taking note that the quadrupole moments depend only on retarded time $u = t - r$, and that the 
angular integrals can be evaluated using the averaging procedure
\be
\langle X \rangle \equiv \frac{1}{4\pi}\, \int d^2 \Og\, X(\thg, \vf) \hs{1} \Rightarrow \hs{1} 
\langle \hat{r}_i \rangle = \langle \hat{r}_{i_1} \hat{r}_{i_2} \hat{r}_{i_3} \rangle = ... = 
\langle \hat{r}_{i_1} ... \hat{r}_{i_{2n+1}} \rangle = 0, 
\label{4.7}
\ee
whilst 
\be
\langle \hat{r}_i \hat{r}_j \rangle = \frac{1}{3}\, \del_{ij}, \hs{2} 
\langle \hat{r}_i \hat{r}_j \hat{r}_k \hat{r}_l \rangle = \frac{1}{15} \lh \del_{ij} \del_{kl} + 
\del_{ik} \del_{jl} + \del_{il} \del_{jk} \rh. 
\label{4.8}
\ee
This results in \ct{Einstein1918}-\ct{Peters1964}
\be
\ba{l}
\dsp{ \frac{dE}{dt} = - \frac{G}{5 c^5}\, \mbox{Tr}\, {\ddd{Q}} {\hs{0.1}^2}, \hs{2} \frac{dP_k}{dt} = 0, }\\
 \\
\dsp{ \frac{dM_k}{dt} = - \frac{2G}{5c^5}\, \ve_{kij} \lh \ddot{Q} \cdot \ddd{Q} \rh_{ij}.  }\\
 \\
\ea
\label{4.9}
\ee
Note that the total flux of linear momentum vanishes by symmetry (in the present approximation) 
as it involves only products of odd numbers of $\hat{\sfr}_i$ integrated over a full spherical surface, 
whereas the integrands of the energy and angular momentum contain even numbers of outward 
spherical unit vectors. 

\section{Generalized newtonian 2-body forces \label{s5}} 

In the following we will apply the results to systems of masses moving under the influence of
mutual newtonian forces, considering two-body systems interacting via a central potential. 
The classical description of such systems simplifies greatly, first as one can 
effectively reduce it to a single-body system by separating off the center-of-mass (CM) motion; 
second as angular momentum conservation implies the relative motion to be confined to a 
two-dimensional plane. Of course, the emission of gravitational radiation introduces limitations 
to these simplifications, but as long as the rate of energy and angular-momentum loss
by the system is small the orbits will change only gradually and one can evaluate the effect 
of gravitational-wave emission in terms of adiabatic changes in the orbital parameters.
In this section we first discuss non-disspiative motion; the effects of gravitational wave 
emission will be analysed afterwards.

Let the bodies have masses $m_1$ and $m_2$ and positions $\bfer_1$ and $\bfer_2$.
To make maximal use of the simplifications we work in the CM frame in which
\[
m_1 \bfer_1 + m_2 \bfer_2 = 0.
\]
In terms of the relative separation vector $\bfer = \bfer_2 - \bfer_1$ the positions w.r.t.\ the 
CM are 
\[
\bfer_1 = - \frac{m_2}{M}\, \bfer, \hs{2} \bfer_2 = \frac{m_1}{M}\, \bfer,
\]
and Newton's third law of motion implies that 
\be
m_1 \ddot{\bfer}_1 = - m_2 \ddot{\bfer}_2 = \mu \ddot{\bfer} = F(r) \hat{\bfer},
\label{5.1}
\ee
where $\mu$ is the reduced mass 
\[
\mu = \frac{m_1 m_2}{m_1 + m_2},
\]
and $F(r)$ is the magnitude of the central force acting on the masses. As usual $r$ and 
$\hat{\bfer}$ represent the modulus and unit direction vector of the separation. In the 
absence of dissipation the energy and angular momentum of the system are conserved. 
In the CM frame these quantities can be written as
\be 
E = \frac{1}{2}\, \mu \dot{\bfer}^2 + V(r), \hs{1} \mbox{such that} \hs{1} F(r) = - \frac{dV}{dr},
\label{5.2}
\ee
and 
\be
\bfL = \mu \bfer \times \dot{\bfer}.
\label{5.3}
\ee
Angular momentum being a conserved vector, the relative motion takes place in the 
plane perpendicular to $\bfL$, which we take to be the equatorial plane $\thg = \pi/2$.
Then 
\be 
\bfer = r \hat{\bfer} = r \lh \cos \vf, \sin \vf, 0 \rh, \hs{2} 
\label{5.4}
\ee
and 
\be
\bfL = \lh 0, 0, \mu \ell \rh, \hs{2} \ell = r^2 \dot{\vf}.
\label{5.5}
\ee
In the following we will always orient the orbit such that the motion is counter-clockwise and 
therefore $\ell \geq 0$. The orbit is represented by the parametrized curve $r(\vf)$ such that 
\be 
\dot{r} = r' \dot{\vf} = \frac{\ell r'}{r^2},
\label{5.6}
\ee
the prime denoting a derivative w.r.t.\ $\vf$. Newton's law of central force (\ref{5.1}) then 
takes the form 
\be
F(r) = \frac{\mu \ell^2}{r^3} \lh \frac{r^{\prime\prime}}{r} - \frac{2r^{\prime\,2}}{r^2} - 1 \rh
 = - \frac{\mu \ell^2}{r^2} \left[ \lh \frac{1}{r} \rh^{\prime\prime} + \frac{1}{r} \right].
\label{5.7}
\ee
This result is tailored to suit Newton's original program of finding the law of force corresponding
to a given orbit \ct{Newton1687}. We will demonstrate it for the particular case of precessing conic 
sections: ellipses, parabolae and hyperbolae; these orbits are parametrized by
\be
r = \frac{\rg}{1 - e \cos n \vf}.
\label{5.8}
\ee
Here $\rg$ is known as the semi-latus rectum; $e$ is the eccentricity: $e = 0$ for circles, 
$0 < e < 1$ for precessing ellipses, $e = 1$  for similar parabolae and $e > 1 $ for hyperbolae. 
Finally the number $n$ determines the rate of precession. For circles this is of course irrelevant. 
For precessing ellipses the apastra occur for
\be
\vf = \frac{2\pi k}{n},  
\label{5.9}
\ee
where $k$ is an integer; thus the apastron shift is $\Del \vf = 2 \pi (1 - n)/n$ per turn. For 
precessing parabolae $n$ determines the angle over which the directrix turns during the 
passage of the two bodies, i.e.\ the asymptotic scattering angle due to precession, also 
measuring 
\be
\Del \vf = \frac{2\pi(1 - n)}{n}.
\label{5.10}
\ee
Similarly for hyperbolae it determines the angle between the incoming and outgoing asymptotes:
\be 
\Del \vf = \vf_{out} - \vf_{in} = \frac{2}{n} \lh \pi - \arccos \frac{1}{e} \rh.
\label{5.11}
\ee
Substitution of the expression (\ref{5.8}) into equation (\ref{5.7}) leads to the result
\be
F(r) = - \frac{\mu n^2 \ell^2}{\rg} \frac{1}{r^2} - \mu (1 - n^2) \ell^2\, \frac{1}{r^3},
\label{5.12}
\ee
the sum of an inverse square and an inverse cube force. Identifying the inverse 
square term with newtonian gravity and introducing an inverse cubic force with strength $\bg \mu$:
\be 
F(r) = - \frac{GM\mu}{r^2} - \frac{\bg \mu}{r^3},
\label{5.13}
\ee
we find
\be
n^2 \ell^2 = GM \rg, \hs{2}  n^2 = \frac{GM\rg}{GM\rg + \bg}.
\label{5.14}
\ee
with $M = m_1 + m_2$ the total mass of the two-body system. Such a force follows from a  
potential
\be
V(r) = - \frac{G\mu M}{r} - \frac{\bg \mu}{2r^2}.
\label{5.15}
\ee
The eccentricity is determined by the radial velocity when the system is at the semi-latus rectum
$\vf = \pi/2n$, $r = \rg$:
\be
\ld \dot{r} \right|_{\vf = \pi/2n} = - \frac{e n \ell}{\rg} = - e\, \sqrt{\frac{GM}{\rg}}.
\label{5.16}
\ee
Evaluating the total energy at the semi-latus rectum and observing it is a constant of motion 
then tells us that 
\be
E = \frac{GM\mu}{2\rg} \lh e^2 - 1 \rh. 
\label{5.17}
\ee
This confirms that for $e^2 < 1$ the orbits are bound, whilst for $e^2 \geq 1$ the orbits are open.
Obviously the total angular momentum is by definition
\be
L_z = \mu \ell = \mu \sqrt{ GM\rg + \bg}.
\label{5.18}
\ee
Note that taking the first-order result for relativistic precession in Schwarzschild  
space-time with innermost circular orbit $R_{isco} = 6GM/c^2$ one gets
\be
n^2 \simeq 1 - \frac{6GM}{c^2 \rg} \hs{1} \Rightarrow \hs{1} \bg = \frac{6 G^2M^2}{c^2} = GM R_{isco}.
\label{5.19}
\ee

\section{Gravitational waves from two-body systems \label{s6}}

In this section and the following we address the emission of gravitational radiation by the two-body 
systems described in section \ref{s5}. As announced we treat this as a form of adiabatic dissipation
changing the orbital parameters $(\rg, e ,n)$ of the system. This applies only to systems in which no 
head-on collisions or mergers involving strong gravity effects take place; these require more powerful 
methods of computation \ct{Blanchet2014}.

To compute the amplitude $h_{ij}$ from equation (\ref{3.12}) for point masses on the quasi-newtonian 
orbits (\ref{5.8}) we must first determine the components of the quadrupole moment and their derivatives. 
For a two-body system in the CM frame they read
\be
\ba{lll}
Q_{ij} & = & \dsp{  m_1 \lh r_{1i} r_{1j} - \frac{1}{3}\, \del_{ij}\, r_1^2 \rh + 
 m_2 \lh r_{2i} r_{2j} - \frac{1}{3}\, \del_{ij}\, r_2^2 \rh }\\
 & & \\
 & = & \dsp{ \mu r^2 \lh \hat{r}_i \hat{r}_j - \frac{1}{3}\, \del_{ij} \rh  
 \equiv \mu r^2 \hat{R}_{ij},  }
\ea
\label{6.1}
\ee
where $\hat{\bfer}$ is the orbital unit vector in the equatorial plane defined in (\ref{5.4}). We explicitly 
factor out the three-tensor array $\hat{\bfR}$ with components $\hat{R}_{ij}$ describing the angular 
dependence of the orbits used in computing the quadrupole moments:
\be
\hat{\bfR} = \frac{1}{2} \left[ \ba{ccc} \cos 2 \vf + \frac{1}{3} & \sin 2 \vf & 0 \\
                                          \sin 2 \vf & - \cos 2 \vf + \frac{1}{3} & 0 \\
                                            0  &  0 & - \frac{2}{3} \ea \right].
\label{6.2}
\ee
Next we want to compute the time derivatives of the quadrupole moment $\bfQ$. For ease of 
computation it is convenient to introduce a set of basic three-tensors in which all our results can 
be expressed:
\be
\bfM = \left[ \ba{ccc} \cos 2\vf & \sin 2 \vf & 0 \\
                             \sin 2 \vf & - \cos 2 \vf & 0 \\
                              0 & 0 & 0 \ea \right], \hs{1}
\bfN = \left[ \ba{ccc} - \sin 2 \vf & \cos 2 \vf & 0 \\
                             \cos 2 \vf & \sin 2 \vf & 0 \\
                              0 & 0 & 0 \ea \right],
\label{6.3}
\ee
and 
\be
\bfI = \left[ \ba{ccc} 1 & 0 & 0 \\
                            0 & 1 & 0 \\
                            0 & 0 & 1 \ea \right], \hs{2} 
\bfJ = \left[ \ba{ccc} 0 & 1 & 0 \\
                            -1 & 0 & 0 \\
                            0 & 0 & 0 \ea \right], \hs{1}
\bfE = \left[ \ba{ccc} \frac{1}{3} & 0 & 0 \\
                             0 & \frac{1}{3} & 0 \\
                             0 & 0 & -\frac{2}{3} \ea \right].                            
\label{6.4}
\ee
They have simple algebraic properties
\be
\ba{l}
\dsp{ \bfE^2 = \frac{2}{9}\, \bfI - \frac{1}{3}\, \bfE, \hs{1} 
\bfM^2 = \bfN^2 = - \bfJ^2 = \frac{2}{3}\, \bfI + \bfE, }\\
 \\
\dsp{ \bfE \cdot \bfM = \bfM \cdot \bfE = \frac{1}{3}\, \bfM, \hs{1} 
         \bfE \cdot \bfN = \bfN \cdot \bfE = \frac{1}{3}\, \bfN, \hs{1}  \bfM \cdot \bfN = - \bfN \cdot \bfM = \bfJ. }
\ea
\label{6.5}
\ee
In addition their derivatives are 
\be
\ba{l}
\dsp{ \frac{d\bfM}{dt} = \frac{2 \ell}{r^2}\, \bfN, \hs{2} \frac{d\bfN}{dt} = - \frac{2\ell}{r^2}\, \bfM, }\\
 \\
\dsp{ \frac{d\bfE}{dt} = \frac{d\bfI}{dt} = \frac{d\bfJ}{dt} = 0. }
\ea
\label{6.6}
\ee
It follows that
\be
\hat{\bfR} = \frac{1}{2} \lh \bfE + \bfM \rh.
\label{6.7}
\ee
Using these results and the ones in appendix \ref{a1} it is now straightforward to establish 
expressions for the quadrupole moment and its derivatives: 
\be
\ba{l}
\dsp{ \bfQ = \frac{\mu r^2}{2}\, \lh \bfE + \bfM \rh,  \hs{2} \dot{\bfQ} = \mu \ell \lh \frac{r'}{r}\, \bfE +
 \frac{r'}{r}\, \bfM + \bfN \rh, }\\
  \\
\dsp{ \ddot{\bfQ} = \frac{\mu \ell^2}{r^2} \left[ \lh \frac{r^{\prime\prime}}{r} - \frac{r^{\prime\,2}}{r^2} \rh \bfE 
 + \lh \frac{r^{\prime\prime}}{r}  - \frac{r^{\prime\,2}}{r^2} - 2 \rh \bfM + \frac{2r'}{r}\, \bfN \right], }\\
 \\
\dsp{ \ddd{\bfQ}\; = \frac{\mu \ell^3}{r^4} \left[ \lh \frac{r^{\prime\prime\prime}}{r} - \frac{5 r^{\prime\prime} r'}{r^2} 
 + \frac{4 r^{\prime\,3}}{r^3} \rh \lh \bfE + \bfM \rh + 4 \lh \frac{r^{\prime\prime}}{r} - \frac{2r^{\prime\,2}}{r^2} 
 - 1 \rh \bfN \right]. }
\ea
\label{6.8}
\ee
More generally we can write for the $n$-th derivative
\be
\bfQ^{(n)} = \frac{\mu \ell^n}{r^{2(n-1)}} \lh Q_E^{(n)}\, \bfE  + Q_M^{(n)}\, \bfM + Q_N^{(n)}\, \bfN \rh,
 \hs{2} n = 0,1,2,3, ...,
\label{6.9}
\ee
where the coefficients $Q^{(n)}_{E, M, N}$ can be read off from the expressions (\ref{6.8}) or
computed by taking still higher derivatives.
These results can now be used to evaluate the amplitude $h_{ij}(\bfx,t)$; the expression 
(\ref{3.12}) for the amplitude is equivalent to 
\be
h_{ij}(\bfx,t) = \frac{\kg}{8\pi \sfr} \left[ \ddot{Q}_{ij} - \hat{\sfr}_i (\ddot{Q} \cdot \hat{\sfr})_j 
 -  \hat{\sfr}_j (\ddot{Q} \cdot \hat{\sfr})_i + \frac{1}{2} \lh \del_{ij} + \hat{\sfr}_i \hat{\sfr}_j \rh 
 \hat{\sfr} \cdot \ddot{Q} \cdot \hat{\sfr} \right]_{u = t - \sfr}.
\label{6.10}
\ee
Note that the direction of the observer is given by the polar unit vector 
\be
\hat{\sfr} = (\sin  \thg \cos \fg, \sin \thg \sin \fg, \cos \thg),
\label{6.11}
\ee
which is distinct from the orbital unit vector $\hat{\bfer}$; then the amplitude in three-tensor notation 
takes the form
\be
\ba{l}
\bfh = \dsp{ \frac{\kg}{8\pi \sfr} \frac{\mu \ell^2}{r^2} \left[ Q^{(2)}_E \bfE + Q^{(2)}_M \bfM 
 + Q^{(2)}_N \bfN \rd }\\
 \\
 \dsp{ \hs{3} -\, \hat{\sfr} \lh Q^{(2)}_E \bfE \cdot \hat{\sfr} + Q^{(2)}_M \bfM \cdot \hat{\sfr}
 + Q^{(2)}_N \bfN \cdot \hat{\sfr} \rh^T - \lh Q^{(2)}_E \bfE \cdot \hat{\sfr} + Q^{(2)}_M \bfM \cdot \hat{\sfr}
 + Q^{(2)}_N \bfN \cdot \hat{\sfr} \rh \hat{\sfr}^T }\\
 \\
 \dsp{ \hs{3} \ld +\, \frac{1}{2} \lh \bfI + \hat{\sfr}\, \hat{\sfr}^T \rh 
 \lh Q^{(2)}_E\, \hat{\sfr} \cdot \bfE \cdot \hat{\sfr} + Q^{(2)}_M\, \hat{\sfr} \cdot \bfM \cdot \hat{\sfr}
 + Q^{(2)}_N\, \hat{\sfr} \cdot \bfN \cdot \hat{\sfr} \rh \right]. }
\ea 
\label{6.12}
\ee
To evaluate this expression use 
\be
\ba{l}
\bfE \cdot \hat{\sfr} = \dsp{ \frac{1}{3} \lh \sin \thg \cos \fg, \sin \thg \sin \fg, - 2 \cos \thg \rh, }\\
 \\
\bfM \cdot \hat{\sfr} = \sin \thg \lh \cos (2\vf - \fg), \sin (2 \vf - \fg), 0 \rh, \\
 \\
\bfN \cdot \hat{\sfr} = \sin \thg \lh - \sin (2 \vf - \fg), \cos (2 \vf - \fg), 0 \rh, 
\ea
\label{6.13}
\ee
and
\be
\hat{\sfr} \cdot \bfE \cdot \hat{\sfr} = \sin^2 \thg - \frac{2}{3}, \hs{1} 
\hat{\sfr} \cdot \bfM \cdot \hat{\sfr} = \sin^2 \thg \cos 2(\fg - \vf), \hs{1}
\hat{\sfr} \cdot \bfN \cdot \hat{\sfr} = \sin^2 \thg \sin 2(\fg - \vf).
\label{6.14}
\ee
The simplest case is that of circular orbits with $r' = 0$ and $\ell = \og r^2$, where 
$\og$ is the constant angular velocity such that $\vf(t) = \og t$. Then
\be
Q^{(2)}_E = Q^{(2)}_N = 0, \hs{1} Q^{(2)}_M = -2,
\label{6.15}
\ee
and 
\be
\bfh = \frac{\kg \mu\, \og^2 r^2}{8\pi \sfr} \left[ - 2 \bfM 
+ 2 \hat{\sfr}\, (\bfM \cdot \hat{\sfr})^T + 2 (\bfM \cdot \hat{\sfr})\, \hat{\sfr}^T 
- \hat{\sfr} \cdot \bfM \cdot \hat{\sfr} \lh \bfI + \hat{\sfr}\, \hat{\sfr}^T \rh \right].
\label{6.16}
\ee
In particular in the equatorial plane $\thg = \pi/2$ and 
\be
\bfh = \frac{\kg \mu\, \og^2 r^2}{16\pi \sfr}\, \cos 2(\fg - \og t) 
 \lh \ba{ccc} 1 - \cos 2 \fg & - \sin 2 \fg & 0 \\
                   - \sin 2 \fg &  1 + \cos 2 \fg & 0 \\
                   0 & 0 & -2 \ea \rh.
\label{6.17}
\ee
whilst along the axis perpendicular to the equatorial plane $\thg = 0$ and
\be
\ba{lll}
\bfh & = & \dsp{ - \frac{\kg \mu\, \og^2 r^2}{4\pi \sfr}\, \bfM }\\
 & & \\
 & = & \dsp{ \frac{\kg \mu\, \og^2 r^2}{4\pi \sfr}   \lh \ba{ccc} \cos 2 \og t & \sin 2 \og t & 0 \\
                             \sin 2 \og t & - \cos 2 \og t & 0 \\
                              0 & 0 & 0 \ea \rh }
\ea
\label{6.18}
\ee
Note that the frequency of the gravitational waves is twice that of the orbital motion, which 
is a direct consequence of their quadrupole nature. 

\section{Radiative energy loss \label{s7}}

The first equation (\ref{4.6}) describes the energy flux of gravitational waves per unit of spherical 
angle as a function of the direction specified by the unit vector $\hat{\sfr}$. Equations (\ref{6.8}) 
specify the quadrupole moments and their derivatives for two-body systems in generalized 
newtonian orbits (\ref{5.8}). To evaluate the differential energy flux these quadrupole moments 
are to be substituted into the energy flux equation. First we compute 
\be
\ba{l}
\dsp{ \left[ \bfQ^{(3)} \right]^2 = \frac{\mu^2 \ell^6}{r^8} \left[ \frac{2}{3} \lh \frac{1}{3}\, Q_E^{(3)\,2} + 
 Q_M^{(3)\,2} + Q_N^{(3)\,2} \rh \bfI + \lh - \frac{1}{3}\, Q_E^{(3)\,2} + Q_M ^{(3)\,2} 
 + Q_N^{(3)\,2} \rh \bfE \rd }\\
 \\
\dsp{\hs{7} \ld +\, \frac{2}{3}\, Q_E^{(3)} Q_M^{(3)}\, \bfM + \frac{2}{3}\, Q_E^{(3)} Q_N^{(3)}\, \bfN \right].}
\ea
\label{7.1}
\ee
It follows that 
\be
\mbox{Tr} \left[ \bfQ^{(3)} \right]^2 = \frac{2\mu^2 \ell^6}{r^8} \lh \frac{1}{3}\, Q_E^{(3)\,2} + 
 Q_M^{(3)\,2} + Q_N^{(3)\,2} \rh,
\label{7.2}
\ee
and 
\be
\ba{l} 
\dsp{ \hat{\sfr} \cdot \left[ \bfQ^{(3)} \right]^2 \cdot \hat{\sfr} = \frac{\mu^2 \ell^6}{r^8} \left[ \frac{4}{9} Q_E^{(3)\,2}
 + \sin^2 \thg \lh - \frac{1}{3}\, Q_E^{(3)\,2} +  Q_M^{(3)\,2} + Q_N^{(3)\,2} \rd \rd }\\
 \\
\dsp{ \hs{9.5} \ld \ld +\, \frac{2}{3} \cos 2 (\fg - \vf)\, Q_E^{(3)} Q_M^{(3)} + 
 \frac{2}{3} \sin 2(\fg - \vf)\, Q_E^{(3)} Q_N^{(3)} \rh \right]. }
\ea
\label{7.3}
\ee
Finally
\be
\hat{\sfr} \cdot \bfQ^{(3)} \cdot \hat{\sfr} = \frac{\mu \ell^3}{r^4} \left[  - \frac{2}{3}\, Q_E^{(3)}
 + \sin^2 \thg \lh Q_E^{(3)} + \cos 2(\fg - \vf)\, Q_M^{(3)} + \sin 2(\fg - \vf)\, Q_N^{(3)} \rh \right].
\label{7.4}
\ee
Inserting the coefficients taken from eq.\ (\ref{6.8}): 
\be
Q_E^{(3)} = Q_M^{(3)} =  \lh \frac{r^{\prime\prime\prime}}{r} - \frac{5 r^{\prime\prime}r'}{r^2} + \frac{4r^{\prime\,3}}{r^3} \rh
 \equiv A, 
\hs{2} Q_N^{(3)} = 4 \lh \frac{r^{\prime\prime}}{r} - \frac{2r^{\prime\,2}}{r^2} - 1 \rh \equiv B,
\label{7.5}
\ee
the general result is
\be
\ba{lll}
\dsp{ \frac{dE}{d^2\Og dt} }& = & \dsp{ - \frac{G\mu^2 \ell^6}{8\pi c^5 r^8} \left[ 
 2 \lh A^2 + B^2 \rh \cos^2 \thg  \frac{}{} \rd}\\
 & & \\
 & & \dsp{ \hs{1} -\, 2\, A^2 \sin^2 \thg \cos 2(\fg- \vf) - 2\, AB \sin^2 \thg \sin 2 (\fg - \vf) }\\
 & & \\
 & & \dsp{ \hs{1} +\, \frac{1}{2}\, \sin^4 \thg \lh A^2 + B^2 + 2\, A^2 \cos 2(\fg - \vf) + 2\, AB \sin 2(\fg - \vf) \rd }\\
 & & \\
 & & \dsp{ \hs{1} \ld \ld +\, \lh A^2 - B^2 \rh \cos^2 2(\fg - \vf) + 
  2\, AB \sin 2(\fg - \vf) \cos 2(\fg - \vf) \rh \frac{}{} \right]. }
\ea
\label{7.6}
\ee
For purely Keplerian orbits this result was derived in \ct{Peters1963}.
Using the results from appendix \ref{a1} for the generalized newtonian orbits (\ref{5.8}) 
the expressions for the quantities $A$ and $B$ take the form 
\be
\ba{l}
\dsp{ A = \frac{n^3 r}{\rg} \sqrt{\lh e^2 -1 \rh \frac{r^2}{\rg^2} + \frac{2r}{\rg} - 1 }, }\\
 \\
\dsp{ B = - \frac{4 n^2 r}{\rg} + 4 \lh n^2 - 1 \rh. }
\ea
\label{7.7}
\ee
The intensity distribution of gravitation radiation emitted by a bound binary system in elliptical 
orbit, precessing and non-precessing, is illustrated for a particular choice of parameters in 
appendix \ref{a2}.

After integrating the result (\ref{7.6}) over all angles the standard result (\ref{4.9}) for the 
total energy loss becomes  
\be
\frac{dE}{dt} = - \frac{2G\mu^2 \ell^6}{15 c^5 r^8}\, \lh4 A^2 + 3 B^2 \rh. 
\label{7.8}
\ee
Substitution of the expressions (\ref{7.7}) then results in
\be
\ba{lll}
\dsp{ \frac{dE}{dt}  }& = & \dsp{ - \frac{8G^4M^3 \mu^2}{15 c^5 n^6 \rg^5} \left[ 
  n^6 \lh e^2 - 1 \rh \frac{\rg^4}{r^4} + 2n^6\, \frac{\rg^5}{r^5} \rd }\\
 & & \\
 & & \dsp{ \hs{4} \ld -\, n^4 \lh n^2 - 12 \rh \frac{\rg^6}{r^6} - 24 n^2 \lh n^2- 1 \rh \frac{\rg^7}{r^7} 
  + 12 (n^2 - 1)^2\, \frac{\rg^8}{r^8} \right]. }
\ea
\label{7.9}
\ee
In the simplest case, that of a circular orbit with $e = 0$, $n = 1$, $r = \rg$ and with angular 
velocity given by
\be
\ell^2 = r^4 \og^2 = GM \rg,
\label{7.10}
\ee
this result reduces to the well-known expression
\be
\frac{dE}{dt} = - \frac{32 G^4 M^3 \mu^2}{5c^5 \rg^5} = 
 - \frac{2}{5} \lh \frac{2GM}{c^2 \rg} \rh^4 \frac{\mu^2 c^3}{M\rg}.
\label{7.11}
\ee
The last result has been cast in terms of the dimensionless compactness parameter $2GM/c^2 \rg$, 
defined as the ratio of the Schwarzschild radius for the combined system and the actual orbital scale 
characterized by $\rg$. For non-precessing orbits for which $n=1$,  $\ell^2 = GM \rg$, the rate of 
energy loss is 
\be
\frac{dE}{dt} = - \frac{1}{30}\, \lh \frac{2GM}{c^2 \rg} \rh^4 \frac{\mu^2 c^3}{M \rg}  
 \left[ \lh e^2 - 1 \rh \frac{\rg^4}{r^4} + 2\, \frac{\rg^5}{r^5} + 11\, \frac{\rg^6}{r^6} \right].
\label{7.12}
\ee
The expression (\ref{7.9}) can also be used to compute the total energy lost by the two-body 
system in a definite period between times $t_1$ and $t_2$, e.g.\ between two periastra 
for bound orbits, or during the total passage of two objects in an open orbit:
\be
\ba{lll}
\Del E & = & \dsp{  \int_{t_1}^{t_2} dt\, \frac{dE}{dt} = 
 \frac{\rg^2}{\ell}\, \int_{\vf_1}^{\vf_2} d \vf\,  \frac{r^2}{\rg^2}\, \frac{dE}{dt} }\\
 & & \\
 & = & \dsp{  \frac{\rg^2}{n\ell}\, \int_{\psi_1}^{\psi_2} d\psi\, \frac{r^2}{\rg^2}\, \frac{dE}{dt}, }
\ea
\label{7.13}
\ee
where we have introduced the integration variable $\psi = n \vf$. 
Now substitute (\ref{7.8}) for the energy change and use 
\[
\frac{\rg}{r} = 1 - e \cos \psi.       
\]
Recalling that $n^2 \ell^2 = GM \rg$ and expanding the integrand transforms the 
expression to 
\be
\ba{l}
\dsp{ \Del E = - \frac{\sqrt{2}}{30n^6} \lh \frac{2GM}{c^2 \rg} \rh^{7/2} \frac{\mu^2 c^2}{M}\,
 \int_{\psi_1}^{\psi_2} d \psi \left[ 12 + n^6 e^2 + e \cos \psi \lh 24 n^2 - 72 - 2n^6 e^2 \rh \rd }\\
 \\
\dsp{ \hs{2} +\, e^2 \cos^2 \psi \lh - n^6 + 12n^4 - 120 n^2 + 180 + n^6 e^2 \rh }\\
 \\
\dsp{ \hs{2} +\, e^3 \cos^3 \psi \lh 2n^6 - 48 n^4 + 240 n^2 - 240 \rh +
 e^4 \cos^4 \psi \lh - n^6 + 72 n^4 - 240 n^2 + 180 \rh }\\
 \\ 
\dsp{ \hs{2} \ld +\, e^5 \cos^5 \psi \lh - 48 n^4 + 120 n^2 - 72 \rh +
 12 (n^2 - 1)^2 e^6 \cos^6 \psi \right]. }
\ea
\label{7.14}
\ee
The adiabatic approximation implies that we treat the parameters $e$ and $n$ in this interval 
as constants; then it is straightforward to perform the integrations. For a bound orbit with 
succesive periastra at $\psi_1 = 0$ and $\psi_2 = 2\pi$ the total energy lost per period to 
gravitational waves is
\be
\ba{l}
\dsp{ \Del E = - \frac{4 \pi \sqrt{2}}{5n^6} \lh \frac{2GM}{c^2 \rg} \rh^{7/2} \frac{\mu^2 c^2}{M}
 \left[ 1 + \frac{e^2}{24} \lh n^6 + 12 n^4 - 120 n^2 + 180 \rh \rd }\\
 \\
\dsp{ \hs{3} \ld +\, \frac{e^4}{96} \lh n^6 + 216 n^4 - 720 n^2 + 540 \rh 
 + \frac{5 e^6}{16} \lh n^2 - 1 \rh^2 \right]. }
\ea
\label{7.15}
\ee
In particular for non-precessing orbits with $n = 1$:
\be
\Del E = - \frac{4 \pi \sqrt{2}}{5} \lh \frac{2GM}{c^2 \rg} \rh^{7/2} \frac{\mu^2 c^2}{M} 
\lh 1 + \frac{73}{24}\, e^2 + \frac{37}{96}\, e^4 \rh.
\label{7.16}
\ee
For the simplest case, a circular orbit with $e = 0$:
\be
\Del E =  - \frac{4 \pi \sqrt{2}}{5} \lh \frac{2GM}{c^2 \rg} \rh^{7/2} \frac{\mu^2 c^2}{M}.
\label{7.17}
\ee
On the other hand, for open orbits with $e \geq 1$ and asymptotic values of the azimuth 
$(\psi_1, \psi_2)$ satisfying
\be
\cos \psi_1 = \cos n \vf_1 = \frac{1}{e}, \hs{1} \sin \psi_1 = \frac{1}{e} \sqrt{e^2 - 1}, 
\hs{1}  \psi_2 = 2 \pi - \psi_1,
\label{7.18}
\ee
the result of the integral (\ref{7.14}) in a somewhat hybrid notation is 
\be
\Del E = - \frac{\sqrt{2}}{ 15 n^6} \lh \frac{2GM}{c^2 \rg} \rh^{7/2} \frac{\mu^2 c^2}{M}\, 
 \sum_{k=0}^6 I_k(n, \psi_1)\, e^k,
\label{7.19}
\ee
with coefficients 
\be
\ba{l}
I_0 = 12  \lh \pi - \psi_1 \rh, \hs{2} I_1 = \lh - 24 n^2 + 72 \rh \sin \psi_1, \\
 \\
\dsp{ I_2 = \frac{1}{2} \lh 3n^6 + 12 n^4 - 120 n^2 + 180 \rh \lh \pi - \psi_1 \rh }\\ 
 \\
\dsp{ \hs{2} +\, \frac{1}{2} \lh n^6 - 12 n^4 + 120 n^2 - 180 \rh \sin \psi_1 \cos \psi_1, }\\
 \\
\dsp{ I_3 = \lh 48 n^4- 240 n^2 + 240 \rh \sin \psi_1 + 
 \frac{1}{3} \lh 2n^6 - 48 n^4 + 240 n^2 - 240 \rh \sin^3 \psi_1,  }\\
 \\
\dsp{ I_4 =  \frac{1}{8} \lh n^6 + 216 n^4 - 720 n^2 + 540 \rh \lh \pi - \psi_1 \rh }\\
 \\
\dsp{ \hs{2} +\, \frac{1}{8} \lh n^6 - 360 n^4 + 1200 n^2 - 900 \rh \sin \psi_1 \cos \psi_1 }\\
 \\
\dsp{ \hs{2} -\, \frac{1}{4} \lh n^6 - 72 n^4 + 240 n^2 - 180 \rh \sin^3 \psi_1 \cos \psi_1,  }\\
 \\
\dsp{ I_5 = \lh 48 n^4 - 120 n^2 + 72 \rh \lh \sin \psi_1 - \frac{2}{3}\, \sin ^3 \psi_1 +  
  \frac{1}{5} \sin^5 \psi_1 \rh,  }\\
 \\
\dsp{ I_6 = 12 \lh n^2 - 1 \rh^2 \left[ \frac{5}{16} \lh \pi - \psi_1 \rh - \cos \psi_1 
 \lh \frac{11}{16}\, \sin \psi_1 - \frac{13}{24}\, \sin^3 \psi_1 + \frac{1}{6}\, \sin^5 \psi_1 \rh \right]. }
\ea
\label{7.20}
\ee
For non-precessing orbits with $n = 1$ the expression simplifies as $I_5 = I_6 = 0$.
The simplest case is the parabolic orbit with $e = 1$, $n= 1$ and $\psi_1 = 0$,
resulting in 
\be
\Del E = - \frac{433 \pi \sqrt{2}}{120} \lh \frac{2GM}{c^2 \rg} \rh^{7/2} \frac{\mu^2 c^2}{M}.
\label{7.21}
\ee
These results are based on the generalized newtonian approximation. Results for scattering in 
the Effective One-Body formalism to all orders in $v/c$ have been obtained in ref.\ \ct{Damour2016}.

\section{Radiative loss of angular momentum  \label{s8}}

The gravitational waves emitted by a system of masses in motion not only carry away energy, 
they also change the system's angular momentum. The last equation (\ref{4.6}) quantifies 
the directional angular momentum loss per unit of time of a non-relativistic system in terms
of the change in the mass quadrupole. In this section we compute the angular momentum 
lost by a quasi-newtonian two-body system as we did for the energy in the previous section.

After substitution of equations (\ref{6.8}), (\ref{6.9}) in the expression (\ref{4.6}) for the 
differential flux of angular momentum we get
\be
\ba{l}
\dsp{ \frac{dM_k}{d^2 \Og dt} = - \frac{G}{4\pi c^5} \frac{\mu^2 \ell^5}{r^6}\, \ve_{kij} 
 \left[ \lh Q^{(2)}_E \bfE + Q^{(2)}_M \bfM + Q^{(2)}_N \bfN \rh \cdot \lh Q^{(3)}_E \bfE 
 + Q^{(3)}_M \bfM + Q^{(3)}_N \bfN \rh_{ij} \rd }\\
 \\
\dsp{ \hs{4.2} 
 - \lh Q^{(2)}_E \bfE \cdot \hat{\sfr} + Q^{(2)}_M \bfM \cdot \hat{\sfr} + Q^{(2)}_N \bfN \cdot \hat{\sfr} \rh_i
 \lh Q^{(3)}_E \bfE \cdot \hat{\sfr} + Q^{(3)}_M \bfM \cdot \hat{\sfr} + Q^{(3)}_N \bfN \cdot \hat{\sfr} \rh_j }\\
 \\
\dsp{ \hs{4.2} +\, \hat{\sfr}_i \lh Q^{(2)}_E \bfE + Q^{(2)}_M \bfM + Q^{(2)}_N \bfN \rh \cdot 
 \lh Q^{(3)}_E \bfE \cdot \hat{\sfr} + Q^{(3)}_M \bfM \cdot \hat{\sfr} + Q^{(3)}_N \bfN \cdot \hat{\sfr} \rh_{j} }\\
 \\
\dsp{ \hs{1} \ld -\, \frac{1}{2}\, \hat{\sfr}_i 
 \lh Q^{(2)}_E \bfE \cdot \hat{\sfr} + Q^{(2)}_M \bfM \cdot \hat{\sfr} + Q^{(2)}_N \bfN \cdot \hat{\sfr} \rh_j
 \lh Q^{(3)}_E \hat{\sfr} \cdot \bfE \cdot \hat{\sfr} + Q^{(3)}_M \hat{\sfr} \cdot \bfM \cdot \hat{\sfr} 
 + Q^{(3)}_N \hat{\sfr} \cdot \bfN \cdot \hat{\sfr} \rh \right] }
\ea
\label{8.1}
\ee
The total loss of angular momentum obtained by integration over all angles as given by 
the result (\ref{4.9}) is
\[
\frac{dM_k}{dt} = - \frac{2G}{5c^5}\, \ve_{kij}\, [ \bfQ^{(2)} \cdot \bfQ^{(3)}]_{ij}.
\]
According to the expansion (\ref{6.9}) and the multiplication rules (\ref{6.5}) the only 
antisymmetric contribution to the product of $\bfQ^{(2)}$ and $\bfQ^{(3)}$ comes from
\[
\bfM \cdot\bfN = - \bfN \cdot \bfM = \bfJ,
\]
which has only a non-vanishing $J_{xy} = - J_{yx} = 1$ component. As the only non-trivial 
component of orbital angular momentum is $M_z$ this is as expected. Using the results of
appendix \ref{a1} it follows that
\be
\ba{l}
\dsp{ \frac{dM_z}{dt} = - \frac{4 G\mu^2 \ell^5 }{5c^5r^6} 
 \lh Q_M^{(2)} Q_N^{(3)} - Q_N^{(2)} Q_M^{(3)} \rh }\\
 \\
\dsp{ \hs{2.5} =\, - \frac{8 G\mu^2 \ell^5 }{5c^5r^6} \left[ n^4 (1- e^2)\, \frac{r^3}{\rg^3} 
 - 2n^2 (n^2 - 1) (1 - e^2)\, \frac{r^2}{\rg^2} + n^2(n^2 + 2)\, \frac{r}{\rg} - 4 (n^2 -1) \right]. }\\
 \\
\ea
\label{8.2}
\ee
For circular orbits with $r = \rg$, $e = 0$ and $n = 1$ this reduces to
\be
\frac{dM_z}{dt} = - \frac{32 G^3 \mu^2 M^2}{5 c^5 \rg^3}\, \sqrt{\frac{GM}{\rg}}
 = - \frac{2\sqrt{2}}{5} \lh \frac{2GM}{c^2 \rg} \rh^{7/2} \frac{\mu^2 c^2}{M},
\label{8.3}
\ee
and for other non-precessing orbits 
\be
\frac{dM_z}{dt} = - \frac{\sqrt{2}}{10} \lh \frac{2GM}{c^2 \rg} \rh^{7/2} \frac{\mu^2 c^2}{M} \left[ (1- e^2)\, \frac{\rg^3}{r^3}
 + 3 \, \frac{\rg^5}{r^5} \right]. 
\label{8.4}
\ee
Following a procedure similar to the treatment of energy we can compute the change in angular 
momentum in a fixed period of time between precessing angles $\psi_{1,2}$:
\be
\ba{l}
\Del M_z = \dsp{ \frac{\rg^2}{n\ell}\, \int_{\psi_1}^{\psi_2} d \psi\, \frac{r^2}{\rg^2}\, \frac{dM_z}{dt} }\\
  \\
\dsp{ \hs{1.5} =\, - \frac{1}{5 n^5} \lh\frac{2GM}{c^2 \rg} \rh^3 \frac{\mu^2 \rg c}{M} \int_{\psi_1}^{\psi_2} d\psi 
  \left[ 4 + e^2 n^2(n^2 -2) \frac{}{} \rd }\\
 \\
 \dsp{ \hs{3} +\, e \cos \psi \lh 6n^2 - 16 - e^2 n^2 (3n^2 - 4) \rh + 
 e^2 \cos^2 \psi \lh n^4 - 16 n^2 + 24 +2 e^2 n^2 (n^2 -1)  \rh }\\
 \\
 \dsp{ \hs{3} \ld +\, e^3 \cos^3 \psi \lh - n^4 + 14 n^2 - 16 \rh 
 - 4 (n^2 - 1) e^4 \cos^4 \psi  \frac{}{}  \right].  }
\ea
\label{8.5}
\ee
It follows that for a bound state the angular momentum lost per period between successive periastra 
$\psi_1 = 0$ and $\psi_2 = 2 \pi$ is
\be
\Del M_z = -\, \frac{8 \pi}{5n^5} \lh \frac{2GM}{c^2 \rg} \rh^3  \frac{\mu^2 \rg c}{M} \left[ 1 
 + \frac{e^2}{8} \lh 3n^4 - 20 n^2 + 24 \rh + \frac{e^4}{8} \lh 2n^2 - 3 \rh \lh n^2 - 1 \rh \right].
\label{8.6}
\ee
For $n = 1$ this becomes:
\be
\Del M_z = -\, \frac{8 \pi}{5n^5} \lh \frac{2GM}{c^2 \rg} \rh^3  \frac{\mu^2 \rg c}{M} \left[ 1 + \frac{7e^2}{8} \right];
\label{8.7}
\ee
for circular motion just take $e = 0$. Next considering open orbits with asymptotic directions 
as in (\ref{7.18}) equation (\ref{8.5}) takes the form
\be
\Del M_z = - \frac{2}{5n^5} \lh \frac{2GM}{c^2 \rg} \rh^3 \frac{\mu^2 \rg c}{M}\, 
 \sum_{k=0}^4 m_k(n,\psi_1)\, e^k,
\label{8.8}
\ee
with coefficients
\be
\ba{l}
m_0 = 4 \lh \pi - \psi_1 \rh, \hs{2} m_1 = \lh - 6 n^2 + 16 \rh \sin \psi_1, \\
 \\
\dsp{ m_2 = \lh \frac{3}{2}\, n^4 - 10 n^2 + 12 \rh \lh \pi - \psi_1 \rh - 
 \lh \frac{1}{2} n^4 - 8 n^2 + 12 \rh  \sin \psi_1 \cos \psi_1, }\\
 \\ 
\dsp{ m_3 = \lh 4n^4 - 18 n^2 + 16 \rh \sin \psi_1 - \frac{1}{3} \lh n^4 - 14 n^2 + 16 \rh \sin^3 \psi_1, }\\
 \\
\dsp{ m_4 = \lh n^2 - 1 \rh \left[ \lh n^2 - \frac{3}{2} \rh \lh \pi - \psi_1 \rh - \lh n^2 - \frac{5}{2} \rh
 \sin \psi_1 \cos \psi_1 - \sin^3 \psi_1 \cos \psi_1 \right]. }
\ea
\label{8.9}
\ee
In particular for {\em parabolic} orbits with $e = n =1$ and $\psi_1 = 0$:
\be
\Del M_z =  - 3 \pi \lh \frac{2GM}{c^2 \rg} \rh^3 \frac{\mu^2 \rg c}{M}.
\label{8.10}
\ee
In ref.\ \ct{Damour1981} a similar result was derived for small-angle scattering in purely 
newtonian gravity with $\bg = 0$.

\section{Evolution of orbits \label{s9}}

The flux of energy and angular momentum carried by gravitational waves as expressed 
by equations (\ref{4.4}) can be determined only if all components of the wave signal are 
known. With present interferometric detectors this is barely possible by combining the 
signals received by at least three instruments at different locations. However, the loss
of energy and angular momentum by sources such as binary star systems is observable 
and allows the gravitational-wave flux to be reconstructed as in the well-known case of 
the binary pulsar systems. Therefore it is of some practical use to evaluate the orbital 
changes due to the emission of gravitational radiation by such systems. Here as in the 
previous sections we consider non-relativistic two-body systems, either in bound orbit 
or on scattering trajectories.

In the adiabatic approximation on which our calculations are based the orbits of two-body 
systems in the CM frame are parametrized by the expression (\ref{5.8}). We take the 
orbital parameters $(\rg, e, n)$ to be slowly changing functions of time; they would be 
constant in the absence of gravitational radiation. According to equations (\ref{5.17}) and 
(\ref{5.18}) the orbital energy and angular momentum are expressed in terms of these 
parameters by
\be
E = \frac{GM\mu}{2\rg} \lh e^2 - 1 \rh, \hs{2} 
L_z = \mu \sqrt{ GM\rg + \bg}.
\label{9.1}
\ee
For comparison with observational data of bound orbits it is sometimes convenient to 
consider the (possibly precessing) semi-major axis of the orbit related to the semi-latus 
rectum by
\be
a = \frac{\rg}{1 - e^2} \hs{1} \Rightarrow \hs{1} E = - \frac{GM\mu}{2a}.
\label{9.2}
\ee
This quantity is also related to the precession parameter by 
\be
\frac{1}{n^2} = 1 + \frac{\bg}{GM\rg} \hs{1} \Rightarrow \hs{1} L_z = \frac{\mu}{n} \sqrt{GM\rg}.
\label{9.3}
\ee
It follows that for bound orbits the orbital parameter changes are related to change in orbital 
energy and angular momentum by
\be
\frac{dE}{dt} = \frac{GM\mu}{2a^2}\, \frac{da}{dt}, \hs{2} 
\frac{dL_z}{dt} = \frac{n\mu}{2} \sqrt{\frac{GM}{\rg}}\, \frac{d\rg}{dt}.
\label{9.4}
\ee
As these parameters are related by (\ref{9.2}) the changes in $\rg$ and in eccentricy $e$
are related as well:
\be
\frac{1}{\rg}\, \frac{d\rg}{dt} = \frac{1}{a}\, \frac{da}{dt} -  \frac{1}{1 - e^2}\, \frac{de^2}{dt}.
\label{9.5}
\ee
Also for constant $\bg$:
\be
\frac{1}{\rg}\, \frac{d\rg}{dt} = \frac{2}{n (1 - n^2)}\, \frac{dn}{dt}.
\label{9.6}
\ee
Now by equating the change in energy and orbital angular momentum to the amount of
energy $\Del E$ and angular momentum $\Del M_z$ carried away by gravitational waves 
we can relate the change in orbital parameters to these parameters themselves. In particular 
according to equations (\ref{7.15})  and (\ref{8.6}) during a period between to succesive 
periastra the orbital parameters change by 
\be
\ba{rll}
\dsp{ \frac{\Del a}{a} }& = & \dsp{ - \frac{\Del E}{E} }\\
 & & \\
 & = & \dsp{ - \frac{16 \pi \sqrt{2}}{5n^6} \frac{\mu}{M} \lh \frac{2GM}{c^2 \rg} \rh^{5/2} \frac{1}{1 - e^2}
 \left[ 1 + \frac{e^2}{24} \lh n^6 + 12 n^4 - 120 n^2 + 180 \rh \rd }\\
 & & \\
 & & \dsp{ \hs{3} \ld +\, \frac{e^4}{96} \lh n^6 + 216 n^4 - 720 n^2 + 540 \rh + \frac{5 e^6}{16} \lh n^2 - 1 \rh^2 \right], }\\ 
 & & \\
\dsp{ \frac{\Del \rg}{\rg} }& = & \dsp{ \frac{2}{n\mu \sqrt{GM\rg}}\, \Del M_z }\\
 & & \\
 & = & \dsp{ -\, \frac{16 \pi \sqrt{2}}{5n^6}  \frac{\mu}{M} \lh \frac{2GM}{c^2 \rg} \rh^{5/2} 
 \left[ 1 + \frac{e^2}{8} \lh 3n^4 - 20 n^2 + 24 \rh + \frac{e^4}{8} \lh 2n^2 - 3 \rh \lh n^2 - 1 \rh \right], }
\ea
\label{9.7}
\ee
Furthermore from these results we can determine the period of the orbit between periastra 
and its evolution. The period itself is
\be
\ba{lll}
T & = & \dsp{ \int_0^{2\pi/n} d\vf\, \frac{dt}{d\vf} = 
  \frac{\rg^2}{n\ell} \int_0^{2\pi} d \psi\, \frac{1}{(1 - e \cos \psi)^2} }\\
 & & \\
 & = & \dsp{ \frac{2\pi}{(1 - e^2)^{3/2}} \frac{\rg^2}{n \ell} = 
  \frac{2\pi  a^{3/2}}{\sqrt{GM}}. }
\ea
\label{9.8}
\ee
This is the appropriate generalization of Kepler's third law for precessing orbits, which holds 
provided the period $T$ is taken to be that between two periastra. From this it follows that 
the rate of change of the period is 
\be
\frac{dT}{dt} = 3 \pi \sqrt{\frac{a}{GM}}\, \frac{da}{dt},
\label{9.9}
\ee
and the relative change per turn is
\be
\frac{\Del T}{T} = \frac{3}{2}\, \frac{\Del a}{a}.
\label{9.10}
\ee
This amounts to a generalization of the Peter-Matthews equation \ct{Peters1963}
\be
\ba{lll}
\dsp{ \frac{dT}{dt} }& \simeq & \dsp{ \frac{\Del T}{T} = 
  - \frac{192 \pi}{5c^5} \frac{G^{5/3} M^{2/3} \mu}{(1 - e^2)^{7/2}} \lh \frac{T}{2\pi} \rh^{-5/3} 
  \left[ \frac{1}{n^6} + \frac{e^2}{24} \lh 1 + \frac{12}{n^2} - \frac{120}{n^4} + \frac{180}{n^6} \rh \rd }\\
 & & \\
 & & \dsp{ \hs{3} \ld +\, \frac{e^4}{96} \lh 1 + \frac{216}{n^2} - \frac{720}{n^4} + \frac{540}{n^6} \rh 
  + \frac{5 e^6}{16 n^6} \lh n^2 - 1 \rh^2 \right]. }
\ea
\label{9.11}
\ee
Next we consider open orbits. These we will characterize in terms of $\rg$ and $e$ 
directly with rates of change determined by (\ref{9.1}) and (\ref{9.4}) 
\be
\frac{1}{e^2 - 1}\, \frac{de^2}{dt} = \frac{1}{\rg}\, \frac{d\rg}{dt} + \frac{1}{E}\, \frac{dE}{dt}.
\label{9.12}
\ee
This results in
\be
\ba{rll}
\dsp{ \frac{d\rg}{dt} \hs {-.5} }& = & \dsp{ \hs {-.5}  -\, \frac{2}{5n^6} \frac{\mu c}{M} \lh \frac{2GM}{c^2\rg} \rh^3 
 \left[ n^4 (1- e^2)\, \frac{\rg^3}{r^3} - 2n^2 (n^2 - 1) (1 - e^2)\, \frac{\rg^4}{r^4} \rd }\\
 & & \\
 & & \dsp{ \hs {-.5} \ld +\, n^2(n^2 + 2)\, \frac{\rg^5}{r^5} - 4 (n^2 -1)\, \frac{\rg^6}{r^6}\, \right], }\\
 & &  \\
\dsp{ \frac{de^2}{dt} \hs{-.5} }& = & \dsp{ \hs{-.5} \frac{1}{60 n^6} \frac{\mu c}{M\rg} 
 \lh \frac{2GM}{c^2 \rg} \rh^3 \left[ 24 n^4 \lh e^2 - 1 \rh^2 \frac{\rg^3}{r^3} \rd }\\
 & & \\
 & & \dsp{ \hs {-.5} -\, n^2 (e^2 - 1) \lh n^4 + 48 (n^2 - 1)(e^2 - 1) \rh \frac{\rg^4}{r^4}  
 - 2n^2 \lh n^4 + 12 (n^2 + 2) (e^2 - 1) \rh \frac{\rg^5}{r^5} }\\
 & & \\
 & & \dsp{ \hs{-.5} \ld +\, \lh n^2(n^2 - 12) + 96 (n^2 - 1)(e^2 - 1) \rh \frac{\rg^6}{r^6} + 24 n^2 \lh n^2 - 1 \rh 
  \frac{\rg^7}{r^7} - 12 \lh n^2 - 1 \rh^2 \frac{\rg^8}{r^8} \right]. }
\ea
\label{9.13}
\ee
The corresponding changes over the complete orbit are 
\be
\frac{\Del \rg}{\rg} = - \frac{4\sqrt{2}}{5n^6}\, \frac{\mu}{M} \lh \frac{2GM}{c^2 \rg} \rh^{5/2} 
 \sum_{k = 0}^4 m_k(n, \psi_1) e^k,
\label{9.14}
\ee
and
\be
\Del e^2 = \lh e^2 - 1 \rh \frac{\Del \rg}{\rg} - \frac{4\sqrt{2}}{15 n^6}\, \frac{\mu}{M} 
 \lh \frac{2GM}{c^2 \rg} \rh^{5/2} \sum_{k=0}^6 I_k(n,\psi_1) e^k.
\label{9.15}
\ee
The total energy change in such an open orbit is given by
\be
\frac{\Del E}{E} = - \frac{4\sqrt{2}}{15 n^6}\,  \frac{\mu}{M} \lh \frac{2 GM}{c^2 \rg} \rh^{5/2}\,
 \frac{\sum_{k=0}^6 (I_k e^k)}{e^2 - 1}.
\label{9.16}
\ee
Finally one can determine for which open orbits the loss of energy by gravitational radiation 
results in a bound orbit, at least in lowest-order approximation. Such a capture process happens 
when the initial energy is positive and the final energy is negative: $ \left| \Del E \right|  > E$. 
From (\ref{9.16}) this requires
\[
\frac{4\sqrt{2}}{15 n^6 (e^2 - 1)}\, \frac{\mu}{M}\, \sum_{k=0}^6 I_k(n, \psi_1) e^k > 
 \lh \frac{c^2 \rg}{2GM} \rh^{5/2}.
\]
As the semi-latus rectum $\rg$ must be greater than the Schwarzschild radius of the system,
the quantity on the left-hand side must be definitely larger than one, and as $\mu < M$ it 
follows that $e^2 - 1$ must be small, i.e.\ the orbit must be close to parabolic. 
\vs{5}

\nit
{\bf Acknowledgement} \\
This paper grew out of a series of lectures by the author at Leiden University in the spring 
of 2018. The support of the Lorentz Foundation throught the Leiden University Fund (LUF) 
is gratefully acknowledged.

\np
\appendix

\section{The transverse traceless gauge \label{a0}}

In this appendix we explain in more detail how starting from an arbitrary solution of the field equations
(\ref{2.3}) for the massless tensor field one can reach the $TT$-gauge (\ref{3.7}) in the far-field region.
We will do this in the hamiltonian formulation in which space- and time components of the fields are 
considered separately. In this formulation the space-components $h_{ij}$ and their conjugate 
momentum fields $\pi_{ij}$ satisfy field equations which are first-order in time derivatives. In contrast
the time components represent auxiliary fields $N = - h_{00}$ and $N_i = h_{0i}$ acting as Lagrage 
multipliers to impose constraints: time-independent field equations restricting the allowed field 
configurations of the space components. The full set of dynamical equations for these fields read 
\be
\ba{lll}
\pi_{ij}  & = & \dot{h}_{ij} - \del_{ij} \dot{h}_{kk} + 2 \del_{ij} \der_k N_k - \der_i N_j - \der_j N_i, \\
 & & \\
\dot{\pi}_{ij} & = &  \Del h_{ij} - \der_i \der_k h_{kj} - \der_j \der_k h_{ki} + \der_i \der_j h_{kk} \\
 & & \\
 & & \hs{2.3} -\, \del_{ij} \lh \Del h_{kk} - \der_k \der_l h_{kl} \rh - \del_{ij} \Del N + \der_i \der_j N + \kg T_{ij}. 
\ea
\label{a0.1}
\ee
The constraints imposed by the auxiliary fields are 
\be
\Del h_{jj} - \der_i \der_j h_{ij} = - \kg T_{00}, \hs{2} \der_j \pi_{ji} = \kg T_{i0}.
\label{a0.2}
\ee
Together these equations are fully equivalent to the covariant field equations (\ref{2.3}). Our
analysis will show that the split in dynamical space- and non-dynamical time components is 
in full agreement with the properties of the causal solutions (\ref{3.1})-(\ref{3.4}). 

As expected the full set of equations (\ref{a0.1}), (\ref{a0.2}) is invariant under local gauge 
transformations which in this formulation take the form 
\be
\ba{ll}
h'_{ij} = h_{ij} + \der_i \xi_j + \der_j \xi_i, & N'_i = N_i + \dot{\xi}_i + \der_i \xi, \\
 & \\
\pi'_{ij} = \pi_{ij} + 2 \del_{ij} \Del \xi - 2\, \der_i \der_j \xi, & N' = N - 2\, \dot{\xi},
\ea
\label{a0.3}
\ee
Clearly the transformations of the auxiliary fields $(N, N_i)$ suffice to remove these 
non-dynamical components by taking 
\be
\dot{\xi} =\frac{1}{2}\, N, \hs{2} \dot{\xi}_i = N_i - \der_i \xi.
\label{a0.4}
\ee
This results in $N' = N'_i = 0$ and 
\be
\ba{lll}
\pi'_{ij}  & = & \dot{h}'_{ij} - \del_{ij} \dot{h}'_{kk}, \\
 & & \\
\dot{\pi}'_{ij} & = &  \Del h'_{ij} - \der_i \der_k h'_{kj} - \der_j \der_k h'_{ki} + \der_i \der_j h'_{kk} 
  - \del_{ij} \lh \Del h'_{kk} - \der_k \der_l h'_{kl} \rh + \kg T_{ij},
\ea
\label{a0.5}
\ee
constrained by
\be
\Del h'_{jj} - \der_i \der_j h'_{ij} = - \kg T_{00}, \hs{2} \der_j \pi'_{ji} = \kg T_{i0}
\label{a0.6}
\ee
Now note that the choice of gauge parameters (\ref{a0.4}) does not fix these transformations 
completely: one can still make residual gauge transformations with parameters $(\xi', \xi'_i)$ 
subject to the conditions
\be
\dot{\xi}' = 0, \hs{2} \dot{\xi}'_i = - \der_i \xi', \hs{2} \ddot{\xi}'_i = 0.
\label{a0.7}
\ee
To see how these can be used, first note that combining the second field equation (\ref{a0.5})
with the first constraint (\ref{a0.6}) results in 
\be
\dot{\pi}'_{jj} = \kg \lh T_{jj} + T_{00} \rh.
\label{a0.8}
\ee
This condition is invariant under the residual gauge transformations, and therefore {\em in empty 
space} where $T_{jj} = T_{00} = 0$ the trace $\pi'_{jj}$ is seen to be constant in time and can be 
removed by a time-independent gauge transformation:
\be
\Del \xi' = \lh - \frac{1}{4}\, \pi'_{jj} \rh_{t=0} \hs{1} \Rightarrow \hs{1} 
\pi^{\prime\prime}_{jj} = \pi'_{jj} + 4 \Del \xi' = 0.
\label{a0.9}
\ee
In view of the first equation (\ref{a0.5}) this also implies that at all times $\dot{h}^{\prime\prime}_{jj} = 0$.
The residual gauge parameters $\xi'_i$ can be used to restrict the field combination
\be
\der_j h_{ji}^{\prime\prime} - \frac{1}{2}\, \der_i h_{jj}^{\prime\prime} = 
 \der_j h'_{ji} - \frac{1}{2}\, \der_i h'_{jj} + \Del \xi'_i. 
\label{a0.10}
\ee
First it can be removed from the initial configuration by taking 
\be
\lh \Del \xi'_i + \der_j h'_{ji} - \frac{1}{2}\, \der_i h'_{jj} \rh_{t = 0} = 0 \hs{1} 
  \Rightarrow \hs{1} \lh \der_j h_{ji}^{\prime\prime} - \frac{1}{2}\, \der_i h_{jj}^{\prime\prime} \rh_{t = 0} = 0.
\label{a0.11}
\ee
In combination with the first constraint (\ref{a0.6}) this implies that in empty space at $t = 0$: 
\be
\lh \Del h_{jj}^{\prime\prime} \frac{}{} \rh_{t=0} = 
   \lh \der_i \der_j h_{ij}^{\prime\prime} \frac{}{} \rh_{t=0} = 0.
\label{a0.12}
\ee
But recall that by the gauge transformation (\ref{a0.9}) we had already achieve that in empty 
space $h_{jj}^{\prime\prime}$ is time-independent; therefore under such conditions the equations 
(\ref{a0.12}) must hold at all times: 
\be
 \Del h_{jj}^{\prime\prime} = \der_i \der_j h_{ij}^{\prime\prime}  = 0.
\label{a0.13}
\ee
Finally one can still make one more residual gauge transformation, with harmonic parameters 
$(\xi^{\prime\prime},\xi_i^{\prime\prime})$ satisfying
\be
\Del \xi_i = 0, \hs{2} \Del \xi = - \der_i \dot{\xi}_i = 0.
\label{a0.14}
\ee
These transformations can be used to remove the trace of the field at $t = 0$,
and therefore at all times: 
\be
h_{jj}^{\prime\prime\prime} = \lh h_{jj}^{\prime\prime\prime} \frac{}{} \rh_{t = 0} = 
  \lh h_{jj}^{\prime\prime} + 2\, \der_i \xi_i^{\prime\prime} \frac{}{} \rh_{t = 0} = 0.
\label{a0.15}
\ee
As the second constraint (\ref{a0.6}) in empty space requires 
\be
\der_j \dot{h}_{ji}^{\prime\prime\prime} = 0, 
\label{a0.17}
\ee
we also find that by combining with (\ref{a0.11}) and (\ref{a0.15})  
\be
\der_j h_{ji}^{\prime\prime\prime} = \lh \der_j h_{ji}^{\prime\prime\prime} \frac{}{} \rh_{t = 0} = 0.
\label{a0.16}
\ee
In conclusion, we have proved that we can find local gauge transformations such that in empty 
space any solution of the field equation can be transformed to the $TT$-gauge 
\[
\der_j h_{ji}^{\prime\prime\prime} = h_{jj}^{\prime\prime\prime} = 0,
\]
by the gauge transformations specified in (\ref{a0.4}), (\ref{a0.9}), (\ref{a0.11}) and (\ref{a0.15}). 

We close this section by noting that the hamiltonian field equations (\ref{a0.1}), (\ref{a0.2}) 
follow directly from the action
\be
S = \int d^4x \lh \dot{h}_{ij} \pi_{ij} - \cH \rh,
\label{a0.18}
\ee
with hamiltonian density
\be
\ba{lll}
\cH & = & \dsp{ \frac{1}{2}\, \pi_{ij}^2 - \frac{1}{4}\, \pi_{jj}^2 + \frac{1}{2} \lh \der_k h_{ij} \rh^2
 - \lh \der_j h_{ji} - \frac{1}{2} \der_i h_{jj} \rh^2 - \frac{1}{4} \lh \der_i h_{jj} \rh^2 }\\
 & & \\
 & & \dsp{  - \kg h_{ij} T_{ij} - 2 N_i \lh \der_j \pi_{ji} - \kg T_{i0} \rh 
  + N \lh \Del h_{jj} - \der_i \der_j h_{ij} + \kg  T_{00} \rh. }
\ea
\label{a0.19}
\ee
In the $TT$-gauge this hamiltonian reduces as expected to the energy density (\ref{4.1}).

\section{Generalized newtonian orbits \label{a1}} 

The generalized newtonian orbits (\ref{5.8}) are parametrized by
\[
r = \frac{\rg}{1 - e \cos n\vf}. 
\]
In our computations we also need the derivatives of this expression, up to the third
derivative. Taking anti-clockwise motion they read
\be
\ba{lll}
\dsp{ \frac{r'}{r} }& = & \dsp{ - n \sqrt{ \lh e^2 - 1 \rh \frac{r^2}{\rg^2} + \frac{2r}{\rg} - 1}, }\\
 & & \\
\dsp{ \frac{r^{\prime\prime}}{r} }& = & \dsp{ n^2 \left[ 2 \lh e^2 - 1 \rh \frac{r^2}{\rg^2} 
 + \frac{3r}{\rg} - 1 \right], }\\
 & & \\
\dsp{ \frac{r^{\prime\prime\prime}}{r} }& = & \dsp{ - n^3 \left[ 6 \lh e^2 - 1 \rh \frac{r^2}{\rg^2} + \frac{6r}{\rg} - 1 \right]
 \sqrt{ \lh e^2 - 1 \rh \frac{r^2}{\rg^2} + \frac{2r}{\rg} - 1}. }
\ea
\label{a1.1}
\ee

\section{Intensity of emission from a binary system \label{a2}}

In this appendix we show an example of the intensity distribution of gravitational-wave 
emission in various directions produced by generalized newtonian binary systems in elliptic 
orbit with eccentricity $e = 0.25$ and precession rates $n = 1$ (newtonian, non-precessing), 
$n = 0.9$ (prograde precession) and $n = 1.1$ (retrograde precession). The intensity 
distribution is represented by the dimensionless quantity
\be
\ba{lll}
Y(\thg,\fg) & = & \dsp{ -128 \pi n^6\, \frac{M^2}{\mu^2} \lh \frac{c^2 \rg}{2GM} \rh^4 
  \frac{\rg\, d(E/Mc^2)}{cdt\, d^2 \Og} }\\ 
  & & \\ 
  & = & \dsp{ \frac{\rg^8}{r^8}  \left[ 2 \lh A^2 + B^2 \rh \cos^2 \thg - 
  2 A^2 \sin^2 \thg \cos 2(\fg- \vf) - 2 AB \sin^2 \thg \sin 2 (\fg - \vf) \frac{}{} \rd }\\
  & & \\
  & & \dsp{ \hs{2} +\, \frac{1}{2}\, \sin^4 \thg \lh A^2 + B^2 + 2 A^2 \cos 2(\fg - \vf) 
  + 2 AB \sin 2(\fg - \vf) \rd }\\
  & & \\
  & & \dsp{ \hs{2} \ld \ld +\, \lh A^2 - B^2 \rh \cos^2 2(\fg - \vf) + 
  2 AB \sin 2(\fg - \vf) \cos 2(\fg - \vf) \rh \frac{}{} \right]. }
\ea
\label{a2.1}
\ee
It is plotted as a function of azimuth $\fg$ for three different polar angles $\thg$: in the equatorial 
plane $\thg = 90^{\circ}$, and in the directions $\thg = 60^{\circ}$ and $\thg = 30^{\circ}$ with
respect to the axis of angular momentum, at three different instants during the orbit where the 
relative orientation of the two masses is  $\vf = 0$, $\vf = 90^{\circ}$ and $\vf = 180^{\circ}$ 
corresponding in the non-precessing case with $n = 1$ to apastron, semi-latus rectum and 
periastron. The same distributions for the same polar angles are also plotted for the case of 
prograde precession with $n= 0.9$, and for retrograde precession with $n = 1.1$.

\np
\bc
\scalebox{0.72}{\includegraphics{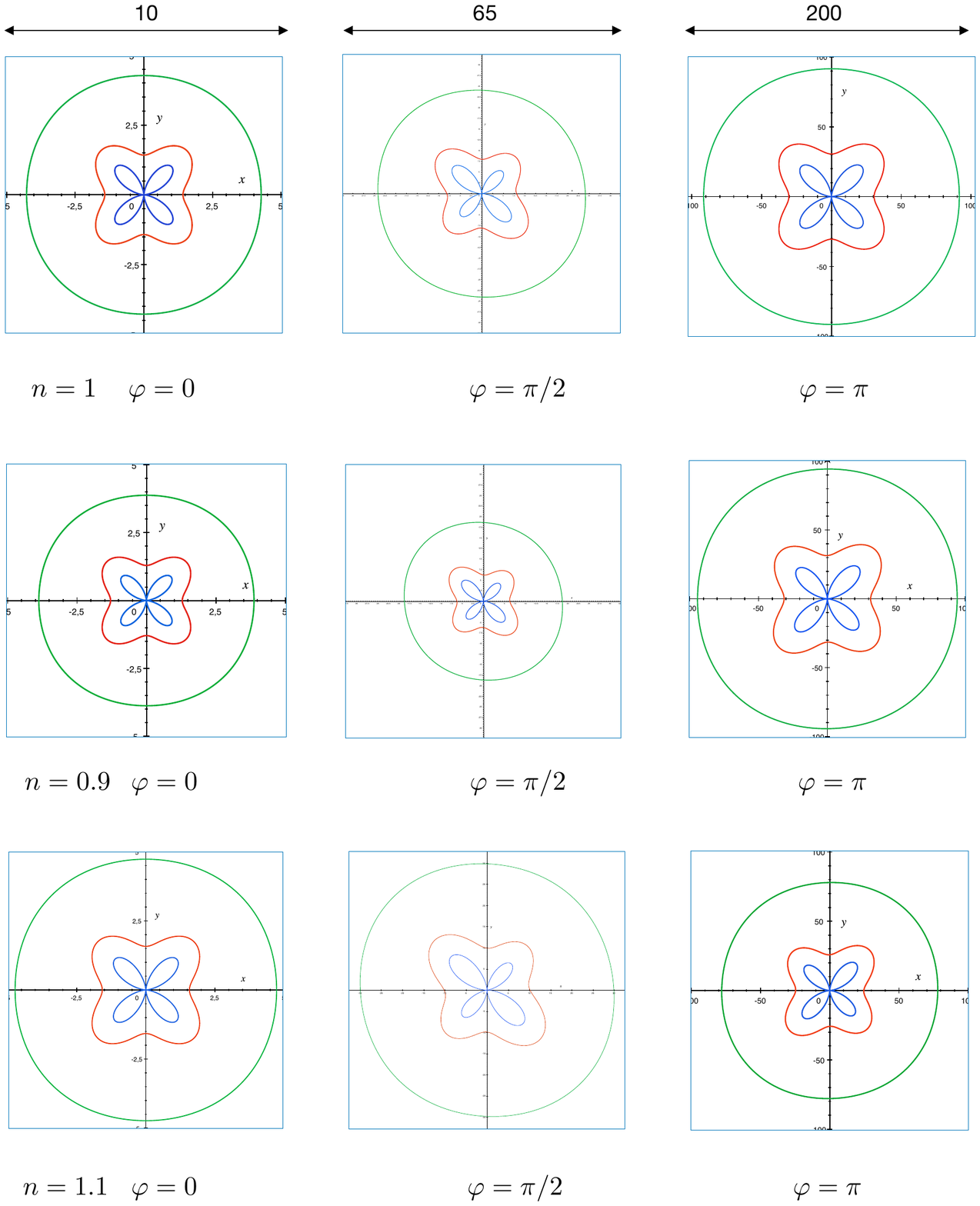}}
\ec
\vs{-20}

\nit
{\footnotesize Fig.\ B1: Intensity patterns of gravitational radiation emitted by a binary system in 
(quasi-)elliptical orbits (characterized by the value of $n$) with eccentricity $e = 0.25$ at three 
different points in the orbit indicated by the values of $\vf$, and in three different directions w.r.t.\ 
the polar axis: $\thg = 90^{\circ}$ (blue inner contour), $\thg = 60^{\circ}$ (red middle contour) and 
$\thg = 30^{\circ}$ (green outer contour). Note that the scales agree in vertical columns, but differ 
from left to right in proportion $10:65:200$.}

\np

\end{document}